\documentclass[aps,showpacs,showkeys,preprintnumbers,amsmath,amssymb]{revtex4}
\usepackage{dcolumn}
\usepackage{graphicx}
\usepackage[utf8]{inputenc}
\def \be {\begin{equation}}
\def \ee {\end{equation}}
\def \bea {\begin{eqnarray}}
\def \eea {\end{eqnarray}}

\begin{document}
\baselineskip=0.8 cm

\title{\bf  Holographic superconductor with dark sector probed by
 entanglement entropy in higher dimensional AdS spacetime }

\author{Weiping Yao $^{a}$, Yifu Cai$^{b}$, and Zehua Tian$^{c,*}$\thanks{yao11a@126.com}}
\affiliation{$^{a}$ School of Physics and Electrical Engineering, Liupanshui Normal University, Liupanshui, Guizhou 553004, P. R. China. Email: yao11a@126.com}

\affiliation{$^{b}$ CAS Key Laboratory for Researches in Galaxies and Cosmology/Department of Astronomy,
School of Astronomy and Space Science, University of Science and Technology of China,
Hefei, Anhui 230026, , P. R. China.  Email: yifucai@ustc.edu.cn}

\affiliation{$^{c}$ School of Physics, Hangzhou Normal University, Hangzhou, Zhejiang 311121, China. Email: tzh@hznu.edu.cn}

\begin{abstract}
\baselineskip=0.6 cm
\begin{center}
{\bf Abstract}
\end{center}
We investigate the holographic entanglement entropy (HEE) with dark sector in a higher-dimensional AdS black hole spacetime including full back reaction, revealing its role as a diagnostic tool for critical phenomena in strongly coupled systems. By analyzing the HEE, we uncover distinct signatures of the metal/superconductor phase transition, demonstrating that the critical temperature is dynamically tuned by both the dark sector coupling strength and the chemical potential ratio between visible and dark sectors. Notably, near the criticality, the HEE exhibits a novel scaling behavior: it grows linearly with the dark sector coupling but displays a nonlinear, accelerated enhancement as the chemical potential ratio between the Maxwell and dark sectors increases.
\end{abstract}

\maketitle

\newpage

\section{Introduction}
The nature of dark sector, which plays a pivotal role in the cosmological evolution of our Universe, remains one of the most profound mysteries in modern physics and cosmology. Compelling evidence for dark sector has been firmly established through multiple independent astrophysical observations, including galaxy rotation curves, gravitational lensing on cluster scales, and temperature fluctuations in the cosmic microwave background \cite{Corbelli1998, Vikhlinin2006, Clowe2006, Planck2015}.
The precision measurements from the Planck satellite mission have quantitatively confirmed that $26.8\%$ of the total energy density of our Universe consists of dark sector \cite{Planck Collaborations2015}, yet its fundamental particle nature and properties remain completely unknown.
In light of this fundamental mystery, unconventional experimental approaches and novel detection techniques are being actively pursued\cite{ADMX2010, Horns2013, Graham2014, Chaudhuri2015, Graham2016, MADMAX2017, CAST2017, Baryakhtar2018, Lawson2019, FUNK2020, Gelmini2020, Gramolin2021, Su2022, Chiles2022, Romanenko2023, Gerard2024}.
Among the most promising avenues is the use of condensed matter systems, particularly superconductors and superconducting devices, which offer unprecedented sensitivity for detecting potential dark sector interactions \cite{Blais2004, Megrant2012, Bertone2018, Hochberg2016, Knapen2017, Knapen2018, Budnik2018, Iwazaki2020, Iwazaki2021, Dixit2021, Wang2022, Chen2023, Acharya2023, Agrawal2024, Guanming2025}.

Recent developments in the anti-de Sitter/conformal field theory (AdS/CFT) correspondence \cite{J.Maldacena1998, EWritten1998, S.S.Gubser1998}
have enable novel studies of superconducting model with considering the dark sector section \cite{Lukasz2014, Lukasz2015,Lukasz2015-1}. These models are constructed in AdS black hole backgrounds incorporating three fundamental fields: (1) a scalar field $\psi$, (2) a Maxwell field $F_{\mu\nu}$ for electromagnetism, and (3) an additional hidden $\mathrm{U}(1)$-gauge field  $B_{\mu\nu}$ representing a component of the dark sector.  The Lagrange density of the Maxwell dark sector is \cite{Vachaspati1991, Achucarro2000}
\begin{equation}\label{L-dark sector}
\mathcal{L}_{DM}=-\frac{1}{4}F_{\mu\nu}F^{\mu\nu}
-[\nabla_{\mu}\psi-\imath A_\mu \psi]^{+}[\nabla^{\mu}\psi-\imath A^{\mu}\psi] \\
-m^2|\psi|^2-\frac{1}{4}B_{\mu\nu}B^{\mu\nu}-\frac{\alpha}{4}F_{\mu\nu}B^{\mu\nu}.
\end{equation}
Note that this dark sector model finds astrophysical support from observations including $511\mathrm{keV}$ gamma rays \cite{Jean2003}, positron excess in galaxies \cite{Chang2008} and muon anomalous magnetic moment \cite{PAMELA2009}. The kinetic mixing parameter $\alpha$ is the a coupling parameter between the Maxwell and dark photon fields. The theoretical studies and experimental observations suggest that the parameter $\alpha$ lies in the range $10^{-3}\leq \alpha \leq10^{-1}$ \cite{Abel2008, Nima2009, Acharya2016, Lees2014, Tilburg2015, Chang2017, Crisler2019, Thomas2022, An2023}. However, in the context of holographic superconductors, the primary interest lies in the boundary CFT, and the physical realism of the dual bulk theory is not essential. To maintain a well-defined dimensionless expectation value for the scalar operator in holographic superconductor model, the factor $\alpha$ is constrained to $|\alpha|<2$ \cite{Katz1940, Aprile2011}. Since the backreaction of matter fields on the metric on the metric will result in a coupling between the electric and energy\cite{HartnollPRL12}, it is essential to consider the backreaction onto the spacetime in this physical models. Considering the influence of the dark sector and  backreaction on the holographic superconductor phase transition, Key findings from Refs.  \cite{Lukasz2014, Lukasz2015,Lukasz2015-1} demonstrate
that the dark sector may significantly modify the critical temperature $T_c$ of phase transitions and the backreaction can bring richer physics in the phase transition.
In this work, we will consider the s-wave holographic superconductor with dark sector including full backreaction in higher dimensional AdS spacetime and explore its phase transition properties.

Furthermore, entanglement entropy has emerged as a powerful diagnostic tool for characterizing strong correlations in quantum systems.
In quantum mechanics, this quantity is formally defined through the von Neumann entropy of a subsystem relative to its complement.
The advent of holographic duality, particularly through the groundbreaking work of Ryu and Takayanagi \cite{Ryu:2006bv, Ryu:2006ef}, has provided a novel computational framework where the entanglement entropy of strongly coupled systems can be determined from their weakly coupled gravitational duals.
This holographic entanglement entropy (HEE) approach has proven particularly valuable in analyzing phase transition properties within holographic superconductor models \cite{Tadashi2012, Cai2012, Xiao-Mei Kuang2014, Dey2014, Aurelio2015, Yan Peng2017, Yunqi Liu2016, David Dudal2018, Fatemeh2019, Hong Guo2019, Yi2020, Sunandan2020, Donald2020, Giorgos2021, Pablo2021, Jeong2022, Yuanceng2023, Dong Wang2023, Hajar2023, Jain2023, Wanhe2024}, where it serves as a sensitive probe of both critical temperatures and transition orders. Recent investigations incorporating dark sector effects have demonstrated HEE's efficacy in studying superconducting phase transitions within four-dimensional anti-de Sitter spacetime \cite{Peng2015}. Building upon these developments and motivated by string-theoretic considerations that naturally require higher-dimensional spacetimes  \cite{Strominger1996, Aharony2000}, we present in this work an extension of holographic entanglement entropy analysis to higher-dimensional AdS black hole backgrounds with coupled dark sectors.

Our aim is to systematically investigate the HEE dynamics in the presence of dark sector interactions and quantitatively analyze their impact on phase transition characteristics. The results demonstrate that HEE serves as a powerful diagnostic tool for both identifying the critical temperature and determining the order parameter of the phase transition. Detailed analysis reveals three key findings regarding HEE behavior near the criticality: (1) The critical HEE exhibits a linear dependence on the dark sector coupling strength $\alpha$; (2) The growth rate shows significant enhancement with increasing chemical potential ratio $\xi/\mu$; (3) Larger geometric width $\ell$ leads to a suppressed critical HEE growth rate, which is exactly opposite to the effect of $\alpha$ and $\xi/\mu$.
These findings establish HEE as a sensitive probe for characterizing how dark sector coupling $\alpha$, chemical potential imbalance $\xi/\mu$, and geometric width $\ell$ collectively influence phase transition properties in holographic systems.

The structure of this work is as follows. In Section \ref{section2}, we present the complete theoretical framework, deriving the equations of motion and boundary
conditions for the holographical superconductor model incorporating dark sector.
Section \ref{section3} systematically examines the behavior of
the scalar condensation operator under the influence of the dark sector. Through detailed analysis of HEE,
Section \ref{section4} investigates the characteristic properties of the phase transition modified
by dark sector.
We conclude with a comprehensive summary of our key findings and their physical implications in the final section.

\section{Equations of motion and boundary conditions for the model}\label{section2}
The action for Einstein-Maxwell scalar dark sector gravity in 5-dimensional
AdS black hole spacetime reads
\begin{eqnarray}\label{action}
S=\int\sqrt{-g} d^{5}x&
\bigg(&\frac{1}{16\pi G_5}[R+\frac{12}{L^2}]-\frac{1}{4}F_{\mu\nu}F^{\mu\nu}
-[\nabla_{\mu}\psi-\imath qA_\mu \psi]^{+}[\nabla^{\mu}\psi-\imath qA^{\mu}\psi]  \nonumber \\
&&-m^2|\psi|^2-\frac{1}{4}B_{\mu\nu}B^{\mu\nu}-\frac{\alpha}{4}F_{\mu\nu}B^{\mu\nu}\bigg),
\end{eqnarray}
where $G_5$ is a 5-dimensional Newton constant in the Einstein gravity on the AdS spacetime,
$R$ is the Ricci scalar, and $L$ is the radius of AdS spacetime
which will be scaled unity in our calculation.
$F_{\mu\nu}$ is the Maxwell field strength tensor.
$m$ and $q$ represent mass and charge
related to the scalar field $\psi$. Since what we are interested in is how the dark sector affects the phase transition physics in the holographic superconductor model,
we shall consider two key parameters: the coupling strength between visible matter field and dark sector field, and the relative proportion between these matter fields.

Without loss of generality in this holographic superconductor model incorporating dark sector, we will take the full back reaction of matter fields on the gravitational sector into consideration. The metric is assumed to be
\begin{eqnarray}\label{BH metric}
ds^2=-f(r)e^{-\chi(r)}dt^{2}+\frac{dr^2}{f(r)}+r^{2}(dx^2+dy^2+d\varsigma^2).
\end{eqnarray}
Correspondingly, the Hawking temperature of the black hole,
which could also be interpreted as the temperature of the CFT, reads
\begin{eqnarray}\label{Hawking temperature}
T_{H}=\frac{f^{\prime}(r_{+})e^{-\chi(r_{+})/2}}{4\pi},
\end{eqnarray}
where $r_{+}$ is the black hole horizon.
We assume that the matter fields posses
only the temporal components which also are real functions of $r$ only.
The ansatz of the matter fields is given by
\begin{eqnarray}
\psi=\psi(r),~~A=\phi(r)dt,~~B_t=\eta(r)dt.
\end{eqnarray}

From the above assumptions, the explicit forms of the field equations are found to be
\begin{eqnarray}
&& \chi^{\prime}+\frac{2}{3}r\left(\psi^{\prime
2}+\frac{q^{2}e^{\chi}\phi^{2}\psi^{2}}{f^{2}}\right)=0,\label{chi}
\\
&& f^{\prime}-2r\left(\frac{2}{L}-\frac{f}{r^2}\right)+\frac{1}{3}r
\left[m^{2}\psi^{2}
+f\left(\psi^{\prime
2}+\frac{q^{2}e^{\chi}\phi^{2}\psi^{2}}{f^{2}}\right)+
\frac{1}{2}e^{\chi}\phi^{\prime2}
+\frac{1}{2}e^{\chi}\eta^{\prime}(\eta^{\prime}+\alpha\phi^{\prime})
\right]=0,\label{f} \nonumber \\
\\
&& \psi^{\prime\prime}+\left(\frac{3}{r}-\frac{\chi^{\prime}}{2}+
\frac{f^\prime}{f}\right)\psi^\prime+\frac{1}{f}\left(\frac{q^{2}e^{\chi}\phi^2}{f}-m^2\right)
\psi=0, \label{psi}
\\
&&\phi^{\prime\prime}+\left(\frac{3}{r}+\frac{\chi^{\prime}}{2}\right)\phi^\prime
-\frac{2q^{2}\psi^{2}}{f}\phi+\frac{\alpha}{2}\left(\eta^{\prime\prime}+
(\frac{3}{r}+\frac{\chi^{\prime}}{2})\eta^\prime\right)=0, \label{phi}
\\
&& \eta^{\prime\prime}+\left(\frac{3}{r}+\frac{\chi^{\prime}}{2}\right)\eta^\prime
+\frac{\alpha}{2}\left(\phi^{\prime\prime}+
(\frac{3}{r}+\frac{\chi^{\prime}}{2})\phi^\prime\right)=0,\label{eta}
\end{eqnarray}
where the prime denotes the derivative with respect to $r$.
To get the solutions of these equations one has to provide appropriate boundary conditions
for this model. Generally the asymptotic behaviors of the fields around the black hole horizon
$r_+$ can be expanded as \cite{Bartlomiej2000}
\begin{eqnarray}\label{Horizon-Boundary}
\nonumber
&&\chi=\chi_0+\chi_1(r-r_+)+..., \label{boundary-chi}\\      \nonumber
&&f=f_0(r-r_+)+f_1(r-r_+)^2+...,\\      \nonumber
&&\psi=\psi_0+\psi_1(r-r_+)+...,\\        \nonumber
&&\phi=\phi_0(r-r_+)+\phi_1(r-r_+)^2+...,\\
&&\eta=\eta_0(r-r_+)+\eta_1(r-r_+)^2+....\label{boundary-eta}
\end{eqnarray}
Near the AdS boundary $(r\rightarrow \infty)$,
the asymptotic behaviors of the solutions yield
\begin{eqnarray}
\chi\rightarrow0\,,\hspace{0.5cm}
f\sim \frac{r^2}{L^2}\,,\hspace{0.5cm}
\psi\sim\frac{\psi_{-}}{r^{\Delta_{-}}}+\frac{\psi_{+}}{r^{\Delta_{+}}}\,,\hspace{0.5cm}
\phi\sim\mu-\frac{\rho}{r^{2}}\,,\hspace{0.5cm}
\eta\sim\xi-\frac{\rho_{d}}{r^{2}},
\label{infinity}
\end{eqnarray}
where $\psi_{-}$ and $\psi_{+}$ denote
the vacuum expectation values $\psi_{-}=\langle\mathcal{O}_{-}\rangle$, $\psi_{+}=\langle\mathcal{O}_{+}\rangle$ of an operator $\mathcal{O}$, respectively,
which are dual to the scalar field according to the AdS/CFT correspondence \cite{HartnollPRL101}.
$\Delta_\pm =2\pm\sqrt{4+m^{2}}$ are the conformal dimensions of
the operators.
$\mu$ and $\rho$ denote the chemical potential and
charge density in the dual field theory, respectively.
$\xi$ and $\rho_d$ are dual to the additional additional hidden $U(1)$ gauge field.

By substituting Eqs. \eqref{Horizon-Boundary} into
the field equations, we can find that there are five independent parameters at the horizon,
i.e., $r_+, \chi_0, \psi_0, \phi_0, \eta_0$. Besides, the useful scaling symmetries in the
the equations of motion  are
\begin{eqnarray}
&& r\rightarrow \lambda r,\qquad (x,y,t)\rightarrow(x,y,t)/\lambda,~~f\rightarrow\lambda^2f
\qquad\phi\rightarrow \lambda\phi,\qquad\eta\rightarrow \lambda\eta, \qquad \label{scaling:r}\\
&& L\rightarrow\lambda L,\qquad r\rightarrow \lambda r,\qquad t\rightarrow \lambda t,\qquad q\rightarrow\lambda^{-1} q \label{scaling:L}.
\end{eqnarray}
Using the above scaling symmetries (\ref{scaling:r}) and (\ref{scaling:L}) we can reset $r_+=1$ and $L=1$.

\section{The condensation of the scalar operator with dark sector}\label{section3}
In this section we will solve the equations of motion for this holographic superconductor model and study the properties of superconducting phase transition through analyzing the scalar operator. Let us begin with the normal phase where $\psi=0$, the solutions of equations of motion can be achieved as the case of the Reissner-Nordstrom-AdS dark sector black object. Specifically,
\begin{eqnarray}
&&\psi=\chi=0, \qquad \phi=\rho(1-1/2r),\qquad \eta=\rho_d(1-1/2r), \\
&&f=r^2-\frac{1}{r^2}\bigg[1+\frac{1}{48}(\rho^2+\alpha \rho^2 \rho_{d}+\rho_{d}^2)\bigg]
+\frac{\rho^2+\alpha \rho^2 \rho_{d}+\rho_{d}^2}{48r^4}.
\end{eqnarray}
In the case of $\psi\neq0$, the equations of motion in superconducting phase are coupled and nonlinear.
To proceed further, we introduce a new variable $z=\frac{r_+}{r}$. For simplicity,  $r_+=1$ is assumed such
that $z\in[1,0]$. In this case $z=1$ represents the black hole horizon and $z=0$ denotes the AdS boundary. Then the equations of motion in $z-$coordinate can be rewritten as
\begin{eqnarray}
&& \chi^{\prime}-\frac{2}{3}z\left(\psi^{\prime
2}-\frac{2q^{2}e^{\chi}\phi^{2}\psi^{2}}{3z^3f^{2}}\right)=0,\label{chi-z}
\\
&& f^{\prime}+\frac{2}{z^3}\left(2-z^2f\right)+\frac{1}{3z^3}
\left[m^{2}\psi^{2}
+f\left(z^4\psi^{\prime
2}+\frac{q^{2}e^{\chi}\phi^{2}\psi^{2}}{f^{2}}\right)+
\frac{z^4}{2}e^{\chi}\phi^{\prime2}
+\frac{z^4}{2}e^{\chi}\eta^{\prime}(\eta^{\prime}+\alpha\phi^{\prime})
\right]=0, \nonumber \\ \label{f-z}
\\
&& \psi^{\prime\prime}-\left(\frac{1}{z}+\frac{\chi^{\prime}}{2}-
\frac{f^\prime}{f}\right)\psi^\prime+\frac{1}{z^2f}\left(m^2-\frac{q^{2}e^{\chi}\phi^2}{z^2f}\right)
\psi=0,  \label{psi-z}
\\
&&\phi^{\prime\prime}-\left(\frac{1}{z}-\frac{\chi^{\prime}}{2}\right)\phi^\prime
-\frac{2q^{2}\psi^{2}}{z^4f}\phi+\frac{\alpha}{2}\left(\eta^{\prime\prime}-
(\frac{1}{z}-\frac{\chi^{\prime}}{2})\eta^\prime\right)=0, \label{phi-z}
\\
&& \eta^{\prime\prime}-\left(\frac{1}{z}-\frac{\chi^{\prime}}{2}\right)\eta^\prime
+\frac{\alpha}{2}\left(\phi^{\prime\prime}-
(\frac{1}{z}-\frac{\chi^{\prime}}{2})\phi^\prime\right)=0.\label{eta-z}
\end{eqnarray}
The equations of motion can be solved numerically by the shooting method.
For specific calculation, note that in five-dimensional spacetime, when $m^2=-\frac{15}{4}>-4$ above Breitenlohner-Freedman bound \cite{Breitenlohner 1982},
the second mode $\psi_+$ is always normalizable. Therefore, in this paper we set $q=2, m^2=-15/4$, and focus on the case where $\psi_-=0$ and  $\psi_+=\langle\mathcal{O}_{+}\rangle$ acts as the vacuum expectation value of the operator $\mathcal{O}_{+}$.
Using the scaling symmetries shown in Eq. (\ref{scaling:r}), we can rescale the relevant quantities as
\begin{equation}\label{scaling-T}
T\rightarrow\lambda T,\qquad
\rho\rightarrow\lambda^3\rho, \qquad
\langle\mathcal{O}_{+}\rangle\rightarrow\lambda^{\frac{5}{2}}\langle\mathcal{O}_{+}\rangle.
\end{equation}

\begin{figure}[htpb]
\centering{
\includegraphics[scale=0.78]{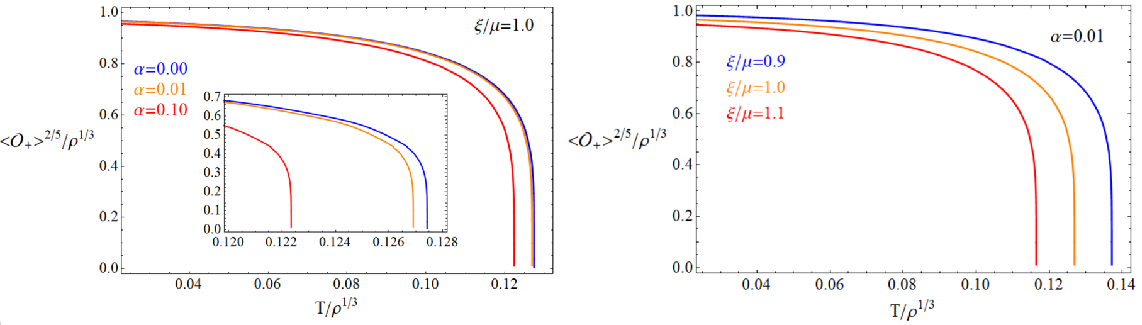}
\caption{\label{condesation} (Color online.) The condensate of operator $\langle \mathcal{O}_{+}\rangle$
versus temperature $T$ for various factors. The left panel shows the superconductor-normal
phase transition with $\frac{\xi}{\mu}=1$ and three lines from top to bottom
correspond to $\alpha=0.00$ (Blue), $\alpha=0.01$ (Orange), $\alpha=0.10$ (red) respectively.
The right panel shows the superconductor-normal
phase transition with $\alpha=0.01$ and three lines from top to bottom
correspond to $\frac{\xi}{\mu}=0.9$ (Blue), $\frac{\xi}{\mu}=1.0$ (Orange), $\frac{\xi}{\mu}=1.1$ (red) respectively.}}
\end{figure}

In figure \ref{condesation}, we plot the behavior of the condensation operator $<\mathcal{O}_{+}>^{\frac{2}{5}}/\rho^{\frac{1}{3}}$ as a function of the temperature, $T/\rho^{\frac{1}{3}}$. With fixed parameters $\alpha$ and $\frac{\xi}{\mu}$,
 We can see from the plots that when the temperature goes below the critical one, $T_c$, the expectation value appears. Which means that the scalar hair forms, identified as a superconductor phase. If $T>T_c$,
the vacuum expectation value of the condensation operator vanishes, meaning there is no scalar field, and thus it is called as the normal phase. At the phase transition point $T=T_c$, the vacuum expectation value of the condensation operator yields $\langle\mathcal{O}_{+}\rangle \varpropto (T_c-T)^{1/2}$,
which means it is a typical second order phase transition\cite{HartnollPRL101, Pan2010}.
It is worth noting that the
solutions go back to the case without dark sector as
$\frac{\xi}{\mu}\rightarrow0$. In this limit, the calculated phase transition temperature $T_c/\rho^{\frac{1}{3}}=0.18999$ is consistent with result in Ref. \cite{yao-5dBI}.
We further observe that as the temperature approaches absolute zero $(T\rightarrow0)$,
all vacuum expectation values of the condensation operator converge to constant values,
which is consistent with the Bardeen-Cooper-Schrieffer (BCS) theory \cite{Bardeen1957}.

How does the dark sector affect the phase transition? In the left panel of figure \ref{condesation}, we take different dark sector coupling factor $\alpha$
and  find that the critical temperature $T_c$ decreases as the factor $\alpha$ increases. It means that the stronger the coupling between the visible field and the dark sector field is, the more difficultly the phase transition occurs. Moreover, when the chemical potential ratio between dark and visible sectors $\xi/\mu$ changes,
we can see from the right panel of figure \ref{condesation} that the bigger $\xi/\mu$ is, the smaller $T_c$ becomes. This means that the scalar field condensation forms harder when the ratio $\frac{\xi}{\mu}$ becomes larger.

\section{Holographic entanglement entropy with dark sector}\label{section4}
In this section,
we are focus on the behavior of the holographic entanglement entropy in this holographic superconductor model.
In the spirit of AdS/CFT correspondence, a simple and elegant proposal to entanglement entropy of the boundary field theory has been presented in Refs. \cite{Ryu:2006bv, Ryu:2006ef}. Specifically,
the holographic entanglement entropy (HEE) for a boundary subregion $A$ is proportional to
the area of a minimal surface $\gamma_A$ in the bulk, whose boundary coincides with
$\partial A$. The HEE of $A$ with its complement is defined as the  ``area law" :
\begin{equation}\label{law}
S_A=\frac{\rm Area(\gamma_\mathcal{A})}{4G_N},
\end{equation}
where $\gamma_A$ is the minimal area surface in the bulk with the same boundary
$\partial A$ of a region A. $G_N$ is Newton constant in the Einstein gravity on the AdS space. From Eq. (\ref{law}), we see that the entanglement entropy depends on the geometry of the
area $A$ which can be chosen arbitrarily. Concretely, we here do the calculations for the areas having a strip geometry $A$ which is given by
\begin{equation}
t=0,\ \ x=x(r), \ \ -\frac{\ell}{2}\leq x \leq \frac{\ell}{2},\ \ -\frac{R}{2}<y<\frac{R}{2}~(R\rightarrow\infty),\ \ -\frac{W}{2}<\zeta<\frac{W}{2}~(W\rightarrow\infty),
\end{equation}
$\ell$ is defined as the size of region $A$. $R$ and $W$ are the regularized lengths in $y$ and $\zeta$ directions, respectively. The holographic surface $\gamma_A$ starts from $r=\frac{1}{\epsilon}$ at $x=\frac{\ell}{2}$, extends into the bulk until it reaches $r=r_*$, then returns back to the AdS boundary $r=\frac{1}{\epsilon}$ at $x=-\frac{\ell}{2}$.
Therefore, the induced metric on the hypersurface $\gamma_A$ can be written as
\begin{equation}
ds^2 =h_{ij}dx^i dx^j=\left (\frac{1}{f(r)}+r^2\left
(\frac{dx}{dr}\right )^2\right )dr^2+r^2 dy^2+r^2d\varsigma^2.
\end{equation}
The area of the minimal surface $\text{Area}(\gamma_A)$ can be expressed as a Lagrangian system
\begin{eqnarray}\label{Area}
\text{Area}(\gamma_A)=2RW\int_{r_*}^{\frac{1}{\epsilon}}
\mathcal{L}_Adr,
\end{eqnarray}
and the Lagrangian $\mathcal{L}_A$ is given by
\begin{eqnarray}\label{Lagrangian}
\mathcal{L}_A=r^2\sqrt{\frac{1}{f(r)}+r^2(dx/dr)^2}dr.
\end{eqnarray}
Minimizing the area of surface $\text{Area}(\gamma_A)$ by the Euler-Lagrange equation,
we can get a identical equation as follows
\begin{equation}\label{d5-minimal}
\frac{r^4(dx/dr)\sqrt{f(r)}}{\sqrt{1+r^2f(r)(dx/dr)^2}}=r_s^3,
\end{equation}
where $r_s$ is a constant.
Considering the surface is smooth at $r=r_*$, we can get $r_s=r_*$
by the condition $dx/dr|_{r=r_*}=0$. By using the variable $z=1/r$,
the HEE can be obtained  as
\begin{eqnarray}\label{HEE}
S_\mathcal{A}=\frac{R W}{2G_5}\int^{z_{*}}_{\epsilon}dz\frac{z_{*}^{3}}{z^{2}}
\frac{1}{\sqrt{(z^{6}_{*}-z^{6})z^{2}f(z)}}
=\frac{R W}{4G_5}\left(\frac{1}{\epsilon^2}+s\right),
\end{eqnarray}
and the width of the belt geometry $\ell$ is
\begin{eqnarray}\label{Length}
\frac{\ell}{2}=\int^{z_{*}}_{\epsilon}dz\frac{z^{3}}{\sqrt{(z^{6}_{*}-z^{6})z^{2}f(z)}}.
\end{eqnarray}
Note that the first term in Eq. (\ref{HEE})
is UV cutoff and represents the ``area law'' \cite{Ryu:2006bv, Ryu:2006ef}.
The second term $s$ is finite term and thus is physically important.
In the exploration of the behavior of the HEE in the holographic superconductor model with
dark sector, we will present our results numerically with dimensionless quantities, $s/ \rho^{\frac{1}{3}}, \ \ T/\rho^{\frac{1}{3}}, \ \ell \rho^{\frac{1}{3}}$.
Specifically, we will analyze how the dark sector coupling factor $\alpha$ and the ratio $\frac{\xi}{\mu}$ affect the HEE, and determinate whether
the HEE can be used to diagnose the phase transition.

\begin{figure}[ht]
\center{
\includegraphics[scale=0.86]{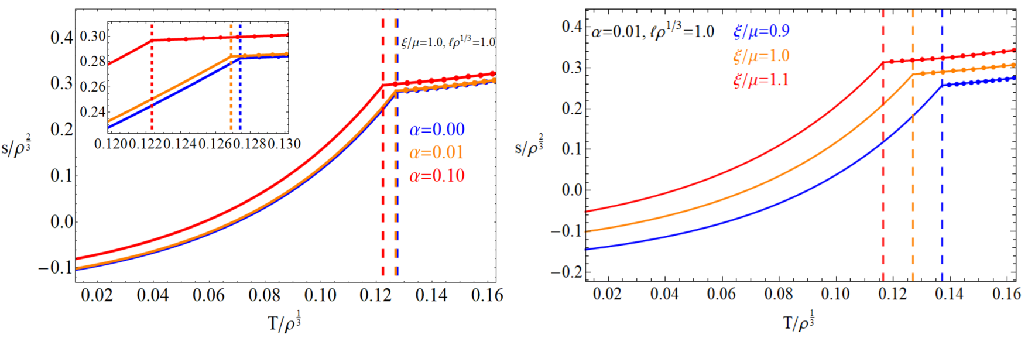}
\caption{ (Color online.)  The HEE s
versus temperature $T$ for various factors. The left panel shows the behavior of
HEE for the constant value of the ratio $\frac{\xi}{\mu}=1$ and the strip width
$\ell \rho^{\frac{1}{3}}=1$. Three lines from top to bottom
correspond to $\alpha=0.00$ (Blue), $\alpha=0.01$ (Orange), $\alpha=0.10$ (red) respectively.
The right panel shows the behavior of
HEE for the constant value of the dark sector coupling constant $\alpha=1$ and the strip width
$\ell \rho^{\frac{1}{3}}=1$. Three lines from top to bottom
correspond to $\frac{\xi}{\mu}=0.9$ (Blue), $\frac{\xi}{\mu}=1.0$ (Orange), $\frac{\xi}{\mu}=1.1$ (red) respectively.} \label{entropy}}
\end{figure}

In figure \ref{entropy}, we show the HEE as a function of temperature $T/\rho^{\frac{1}{3}}$ for different
values of dark sector coupling factor $\alpha$ and the ratio $\frac{\xi}{\mu}$.
Note that the vertical dashed lines represent the critical temperature of the normal-superconductor phase transition,
and the dot-dashed lines and the solid ones denote the normal phase and the superconductor cases, respectively.
We can see from the plots that the HEE decreases as temperature drops. Notably, the HEE in the superconductor phase is consistently lower than in the normal phase.
Moreover, as temperature drops, the HEE in the superconductor regime declines more sharply compared to that in the normal phase. This behavior persists across varying values of the parameters $\alpha$ and $\frac{\xi}{\mu}$. The observed behavior of the HEE is due to the fact that the metal phase can be thought of as the one filled with free charge carriers, such as electrons. At the critical temperature $T_c$, the condensate turns on and the free charge carriers are continuously condensed to Cooper pairs as temperature decreases.  This pairing mechanism reduces the available degrees of freedom in the condensed phase, thereby suppressing the HEE. It's worth noting that the HEE at $T_c$ is continuous, while its first derivative with respect to the temperature is discontinuous. Which shows that the phase transition is the second order.
Therefore, in this regard, one can determinate the type of phase transition from the behavior of the HEE.

When we fixed the parameter $\xi/\mu$ and $\ell\rho^{1/3}$, we can see from the left panel in figure \ref{entropy} that the critical temperature $T_c$ decreases with the increase of the dark sector coupling factor $\alpha$. In the right panel of figure \ref{entropy}, we can also find that the bigger ratio $\frac{\xi}{\mu}$ leads to smaller critical temperature $T_c$.
This suggests that both the stronger coupling factor $\alpha$ and bigger ratio $\frac{\xi}{\mu}$ will cause scalar field condensation harder.
In particular, in the tables \ref{Tc-a} and \ref{Tc-emu}, we obtain the critical temperature from the analysis of the the HEE, and compare them with that from
analyzing the behaviors of the scalar operator. We observe that the critical temperature obtained from these two different physical scenarios are completely consistent. That is to say, the HEE can also be used to probe the point of the phase transition.

\begin{table}[ht]
\caption{\label{Tc-a}  The critical temperature $T_c$ obtained by the condensation of the scalar operator $ <\mathcal{O}_{+}>$ and from behaviors of the HEE with different values of $\alpha$ for $\frac{\xi}{\mu}=1, \ell \rho^{\frac{1}{3}}=1$}
\begin{tabular}{c c c c c c c c}
         \hline
$\alpha$ & 0 & 0.2 & 0.4 & 0.6 & 0.8 & 1.0
        \\
        \hline
~~$<\mathcal{O}_{+}>$~~&~~$0.12743$~~&~~$0.11745$~~&~~$0.10823$~~&~~
$0.099733$~~&~~$0.09191$~~&~~$0.08471$~~~
          \\
~~$HEE$~~&~~$0.12743$~~&~~$0.11745$~~&~~$0.10823$~~&~~
$0.099733$~~&~~$0.09191$~~&~~$0.08471$~~
          \\
        \hline
\end{tabular}
\end{table}

\begin{table}[ht]
\caption{\label{Tc-emu} The critical temperature $T_c$ obtained by the condensation of the scalar operator $ <\mathcal{O}_{+}>$ and from behaviors of the HEE with different values of $\frac{\xi}{\mu}$ for $\alpha=0.01, \ell \rho^{\frac{1}{3}}=1$
}
\begin{tabular}{c c c c c c c}
         \hline
$\xi/\mu$ & 0 & 0.2 & 0.4 & 0.6 & 0.8 & 1.0
        \\
        \hline
~~$<\mathcal{O}_{+}>$~~&~~$0.18999$~~&~~$0.18691$~~&~~$0.17817$~~&~~
$0.16449$~~&~~$0.14694$~~&~~$0.12691$~~
          \\
~~$HEE$~~&~~$0.18999$~~&~~$0.18691$~~&~~$0.17817$~~&~~
$0.16449$~~&~~$0.14694$~~&~~$0.12691$~~
          \\
        \hline
\end{tabular}
\end{table}

\begin{figure}[ht]
\center{
\includegraphics[scale=0.58]{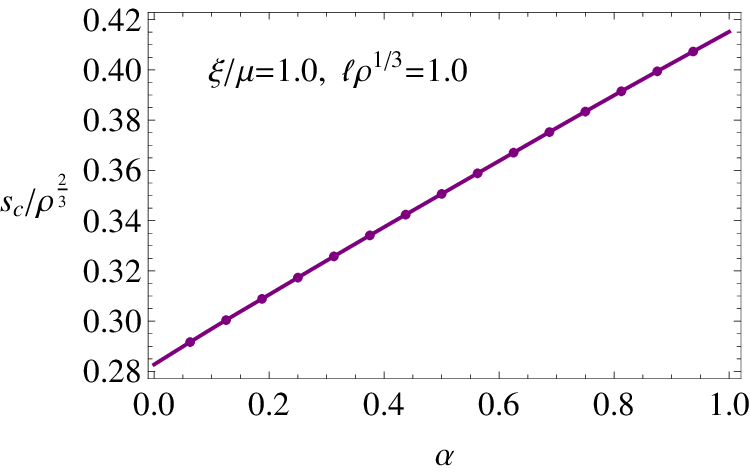}
\includegraphics[scale=0.58]{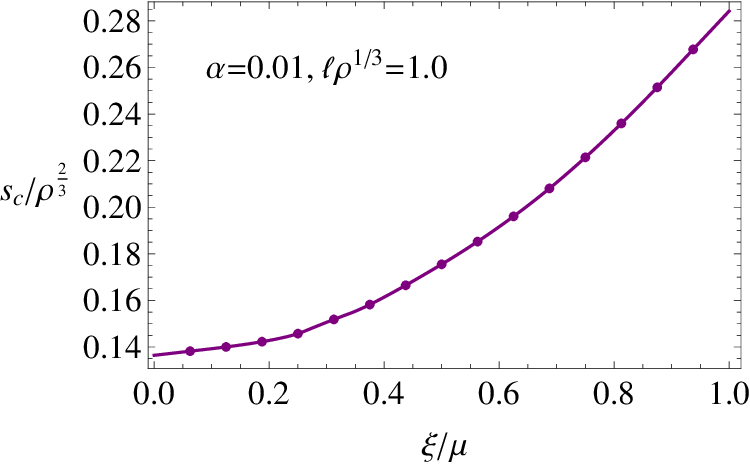}
\caption{ (Color online.) The behaviors of the critical HEE $s_C$ at the phase transition point for various factors $\alpha$ and $\frac{\xi}{\mu}$.} \label{critical-sC}}
\end{figure}

To gain deeper insight into how dark sector influences phase transition properties,
we in the following focus on the behavior of the critical HEE $s_C$ with respect to
 the factors $\alpha$ and $\frac{\xi}{\mu}$ respectively.
It is shown in figure \ref{critical-sC} that the critical HEE $s_C$ at the phase transition point increase as the factors $\alpha$ and $\frac{\xi}{\mu}$ become bigger. However, the critical HEE increases linearly with the increase of the coupling factor $\alpha$, while it
increases nonlinearly with the increase of the ratio $\frac{\xi}{\mu}$. These distinctly different behaviors can be used as a criteria to determine
 which, the coupling factor $\alpha$ or the ratio $\frac{\xi}{\mu}$, induces the phase transition.

\begin{figure}[ht]
\center{
\includegraphics[scale=0.35]{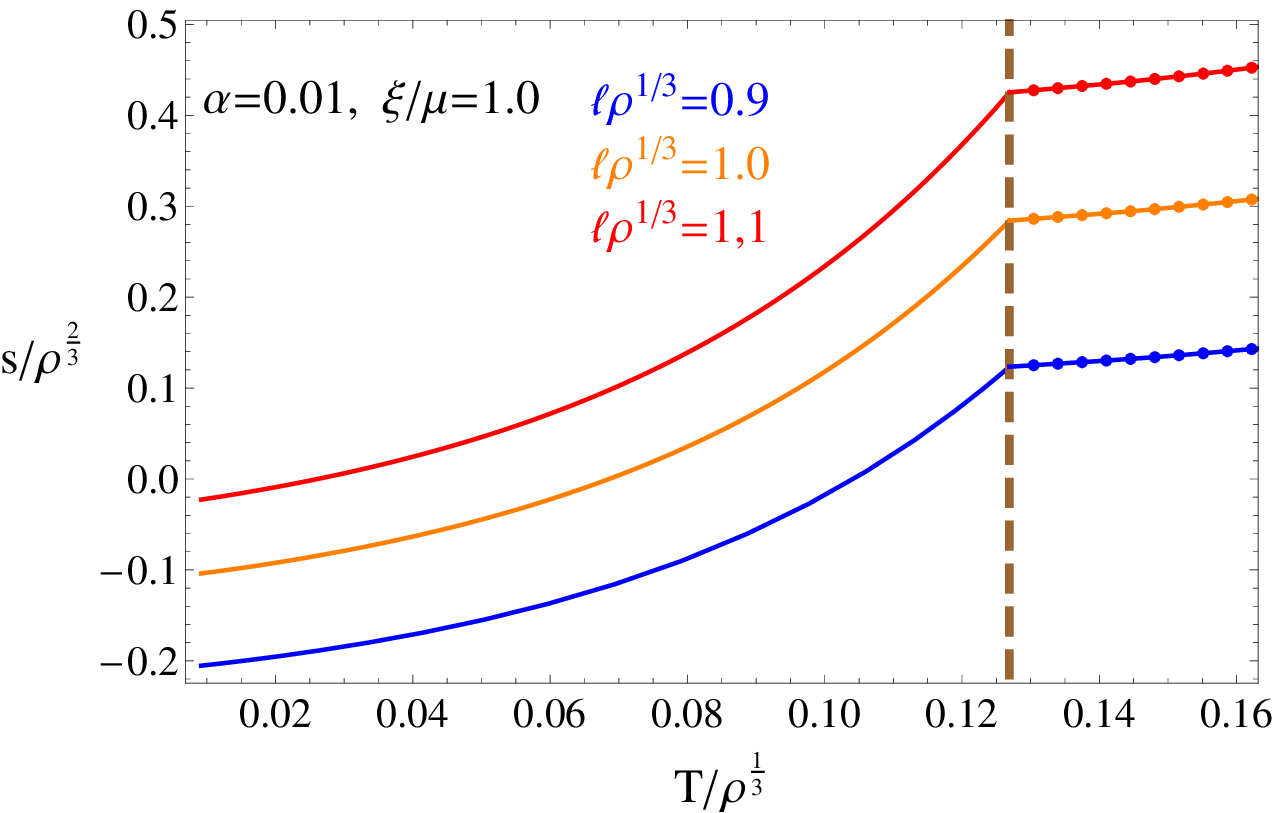}
\includegraphics[scale=0.47]{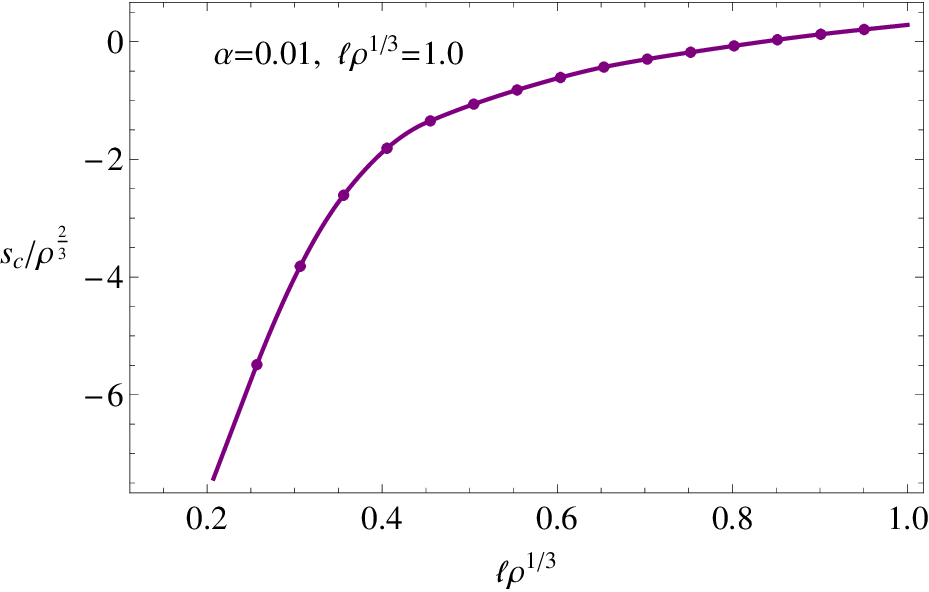}
\caption{ (Color online.) The behaviors of the HEE versus temperature $T$ for
different width $\ell$ with $\alpha=0.01,\frac{\xi}{\mu}=1$ (left plot).
Three lines from top to bottom
correspond to $\ell\rho^{\frac{1}{3}}=0.9$ (Blue), $\ell\rho^{\frac{1}{3}}=1.0$ (Orange), $\ell\rho^{\frac{1}{3}}=1.1$ (red) respectively. The right plot shows the corresponding critical HEE $s_C$ at the phase transition point versus the width $\ell$.} \label{critical-L}}
\end{figure}

In figure \ref{critical-L}, we show how the belt width $\ell$ affects the HEE. We can find that when we fix the coupling factor $\alpha$ and the ratio
$\xi/\mu$, the critical temperature $T_c$ is independent of the width $\ell$, and it remains constant with various $\ell$.
This suggests that the belt width of the geometry have no effect on the phase transition point. When we fix the temperature $T$,
it is found that the HEE becomes bigger as the width $\ell$ increase. At the phase transition point, we can see from the right panel of figure \ref{critical-L}
that the HEE $s_C$ increases slowly as the width $\ell$ grows, which behaves quite differently from the cases when increasing the parameters $\alpha$ and $\frac{\xi}{\mu}$.

\section{Conclusion}
The search for dark sector signatures in astrophysical phenomena remains a pivotal challenge in modern physics. In this work, we employ the HEE as a diagnostic tool to investigate phase transitions in a higher-dimensional AdS spacetime incorporating dark sector.
Taking the full back reaction
of the matter field sector on the gravitational background into account,
we found that the critical temperature values derived from HEE behavior align with the phase transition points identified through the scalar operator.
At the phase transition point, the jump of the slope of HEE indicates the phase transition is
second order. These findings underscore that the HEE serves as an effective probe for identifying the critical temperature and characterizing the order of the phase transition in this physical holographic superconductor model. Increasing the dark sector coupling factor $\alpha$ and the ratio of the chemical potential between the visible and dark sectors $\frac{\xi}{\mu}$ reduces the critical temperature, making condensation more difficult.
 Moreover, based on the probe of HEE, it is worth to study the critical behavior
of the HEE versus the dark sector coupling factor $\alpha$ and the ratio $\frac{\xi}{\mu}$.
In figure \ref{critical-sC}, we observed that the HEE at the phase transition point
increases linearly with the enhancement of the dark sector coupling factor. However, the critical HEE exhibits nonlinearly increase as the ratio of the chemical potential between the visible and dark sectors grows. We also examine the influence of geometric width on the critical HEE, observing that larger widths result in a slower increase of HEE at the phase transition point. This behavior is different from the cases of the critical HEE versus the parameters $\alpha$ and $\frac{\xi}{\mu}$.

\section{Acknowledgments}
We sincerely thank Prof. Jian Li for his constructive suggestions and support.
This work of W. Y. was supported by the National Natural Science Foundation of China under Grant No. 11665015; Guizhou Provincial Science and Technology Planning Project of
China under Grant No. qiankehejichu-ZK[2021]yiban026; the Guizhou Province ordinary institutions of higher learning top science and technology talents Support plan of China
(qianjiaoheKY[2019]062); the project of the Discipline Team for the Master's Program Construction at Liupanshui Normal University under Grant No. LPSSY2025XKTD09; Liupanshui Science and Technology Development Project(NO.52020-2024-PT-01);
Y. C. acknowledges the National Key $R\&D$
Research Funds for Central Universities, CSC Innovation
Talent Funds, USTC Fellowship for International Cooperation, USTC Research Funds of the Double First-Class
Initiative, CAS project for young scientists in basic research (YSBR-006);
Z. T. was supported by the scientific
research start-up funds of Hangzhou Normal University:
4245C50224204016; the National Natural Science Foundation of China under Grant No. 12575050.

\end{document}